\documentclass[preprint,preprintnumbers,amsmath,amssymb]{revtex4}
\usepackage{graphicx,color}

\begin{document}



\title{\itshape Influence of the Anisometry of Magnetic Particles on the Isotropic-Nematic Phase Transition}

\author{V. Gdovinov\'a$^{\rm a}$, N. Toma\v{s}ovi\v{c}ov\'a$^{\rm a}$$^{\ast}$\thanks{$^\ast$Corresponding author. Email:
nhudak@saske.sk \vspace{6pt}}, N. \'Eber$^{\rm b}$, T.
T\'oth-Katona$^{\rm b}$, V. Z\'avi\v{s}ov\'a$^{\rm a}$, M.
Timko$^{\rm a}$ and P. Kop\v{c}ansk\'y$^{\rm a}$\\\vspace{6pt}
$^{a}${\em{Institute of Experimental Physics, Slovak Academy of
Sciences, Watsonov\'a 47, 04001 Ko\v{s}ice, Slovakia}};
$^{b}${\em{Institute for Solid State Physics and Optics,  Wigner
Research Centre  for Physics,  Hungarian Academy of Sciences, H-1525
Budapest, P.O.Box 49, Hungary}}\\
}

\begin{abstract}
The influence of the shape anisotropy of magnetic particles on the
isotropic-nematic phase transition was studied in ferronematics
based on the nematic liquid crystal
4-(trans-4-n-hexylcyclohexyl)-isothiocyanato-benzene (6CHBT). The
liquid crystal was doped with spherical or rod-like magnetic
particles of different size and volume concentrations. The phase
transition from isotropic to nematic phase was observed by
polarizing microscope as well as by capacitance measurements. The
 influence of the concentration and the shape
anisotropy of the magnetic particles on the isotropic-nematic phase
transition in liquid crystal is demonstrated. The results are in a good agreement
with recent theoretical predictions.


\end{abstract}

\maketitle

\section{Introduction}

Doping liquid crystals (LCs) with nanoparticles (NPs) in low volume
concentrations has been shown as a promising method to modify the
properties of liquid crystals. The presence of nanoparticles in
liquid crystal changes the properties of the mesophase and/or
introduces some new features of the composite mixtures. After
introducing the idea  theoretically  by Brochard and de Gennes
\cite{BG}  several reports have shown  that the doping of liquid
crystals with magnetic nanoparticles can either increase or decrease
the critical field of the magnetic Fr\'eedericksz transition
\cite{PK1,PK2,Podoliak1,Chapter2012}, depending on the host-guest
combination. It has also been proven \cite{Podoliak2,NT1} that
ferronematics respond to low magnetic fields  (below 0.1~T) due to
doping with magnetic particles. Furthermore, the theoretically
predicted magnetic field-induced isotropic-nematic phase
transition \cite{Lelidis1993} was observed in calamitic liquid crystals
at relatively low magnetic fields (below 12~T) when they are doped with
magnetic nanoparticles \cite{PK3}.

Properties of magnetic nanoparticles significantly depend on their
size, shape and structure. Therefore, the
properties of ferronematics composed of them are also expected to be
sensitive to the same factors. As an example, in our previous work
\cite{PK1} it was shown that doping the liquid crystal
4-(trans-4-n-hexylcyclohexyl)-isothiocyanato-benzene (6CHBT) with
spherical magnetic particles results in soft anchoring
($\omega$~$\sim$~1), while in the case of doping with rod-like
magnetic particles rigid anchoring ($\omega$~$\sim$~10$^{4}$) was
obtained at the LC-NP interface, even though both kinds of magnetic
particles were coated with the same surfactant (oleic acid). Here,
the parameter  $\omega$~=~$W R/K$ is the ratio of the interfacial
(anchoring) to the bulk (elastic) energies; $W$ is the anchoring
energy density, $R$ is the reference size of the NPs \cite{Burylov1995} which in case
of cylindrical particles coincides with their mean radius $D/2$ \cite{Burylov1995,Burylov1994},
and $K$ is the corresponding elastic modulus of the host LC.

Recently, a mean-field theory has been developed by Gorkunov and
Osipov \cite{GO} to describe the influence of embedded nanoparticles
on the orientational order and on the isotropic-nematic phase
transition of the host liquid crystal. They considered  relatively
large, uniaxial NPs of oblate or prolate shape, where the
interaction between the LC molecules and the NPs may be described by
an anisotropic surface potential depending on the coupling between
the surface normal of the particle and the long molecular axis of
the host. It has been shown that spherically isotropic nanoparticles
effectively dilute the liquid crystal medium and decrease the
isotropic-nematic transition temperature. On the contrary,
anisotropic nanoparticles become aligned by the nematic host and,
reciprocally, improve the liquid crystal alignment, thus extending
the nematic order to higher temperatures.

 This theory has inspired
us to perform measurements with the nematic liquid crystal
6CHBT doped with spherical or rod-shaped magnetic nanoparticles. The
present work is devoted to an experimental study of how the shape, as
well as the volume concentration of magnetic nanoparticles affect
the temperature of the isotropic-nematic phase transition. The
obtained results indicate a significant effect in accordance with
the theoretical expectations described in \cite{GO}.

\section{Experiment}

\begin{figure}
\begin{center}
{\bf (a)}\includegraphics[width=20pc]{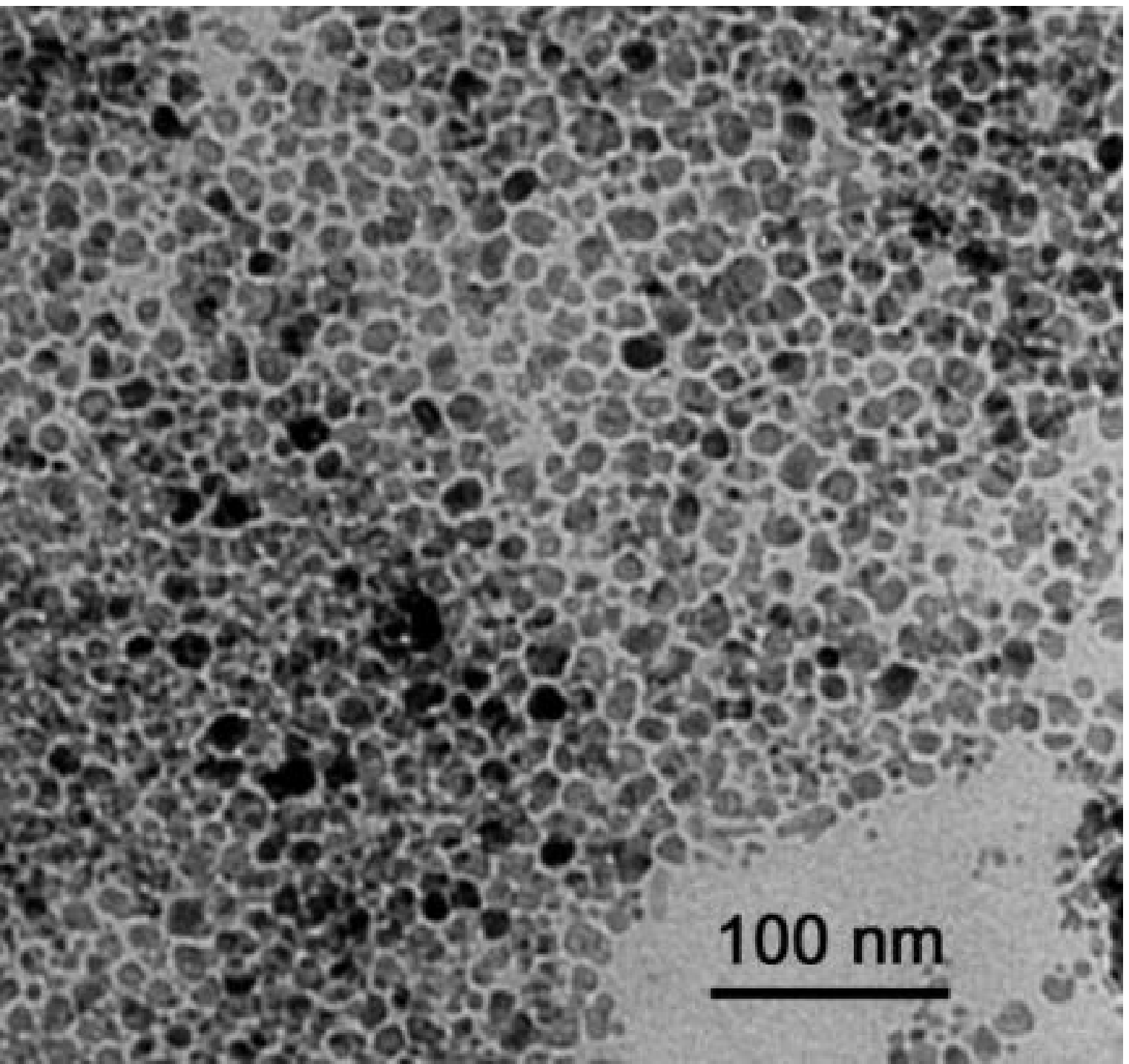} \\
{\bf (b)}\includegraphics[width=20pc]{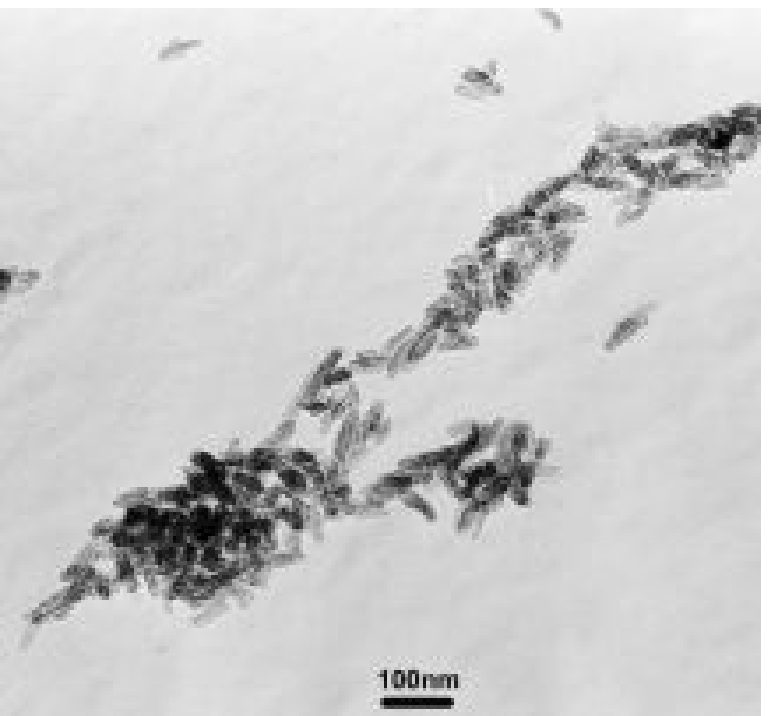}\\
{\bf (c)}\includegraphics[width=20pc]{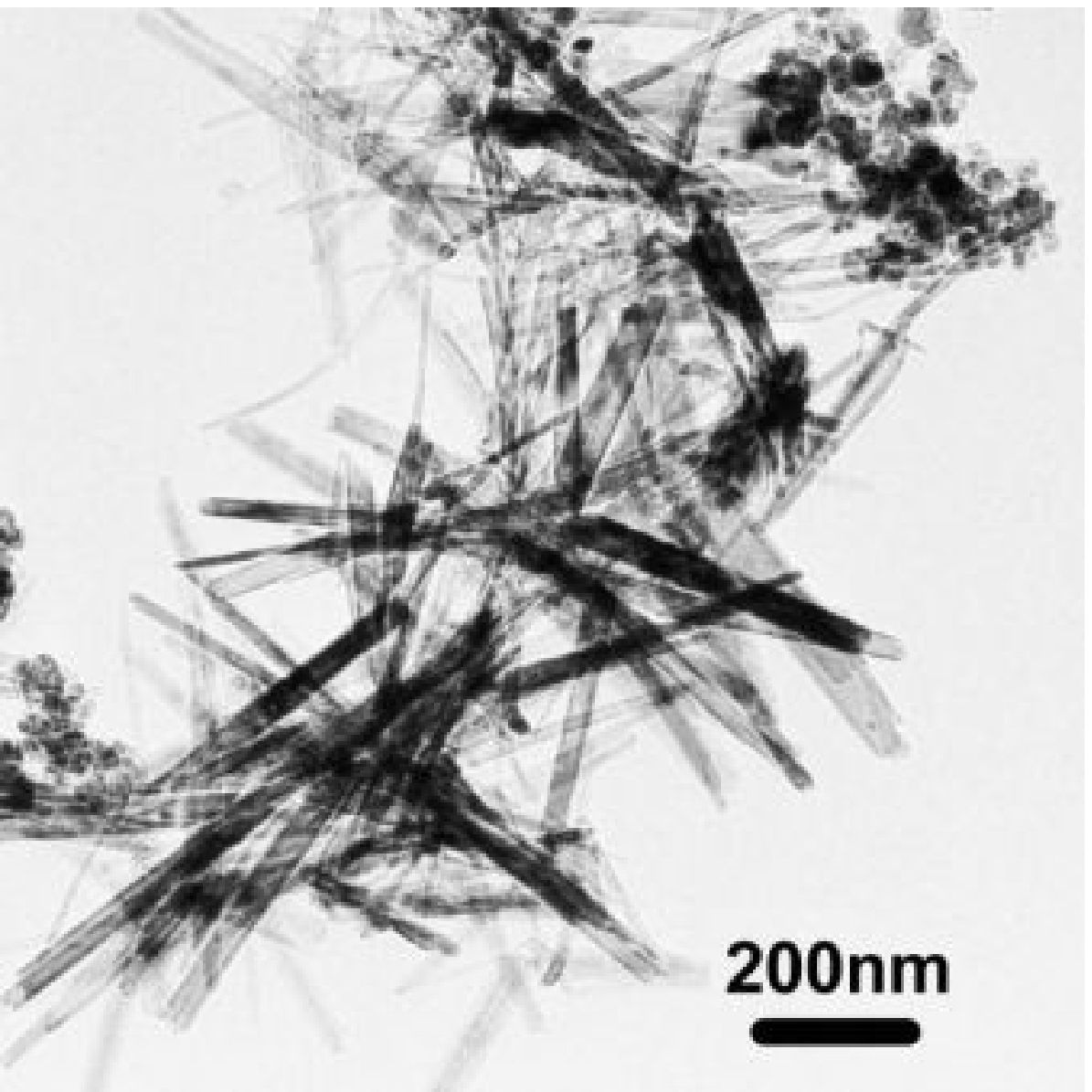} \caption{TEM images of
(a) spherical, (b) short and (c) long rod-like magnetic
nanoparticles.} \label{TEM}
\end{center}
\end{figure}

The spherical magnetic NPs were prepared by the co-precipitation method
described in \cite{PK1}. The rod-like iron oxide NPs were
synthesized through smooth decomposition of urea \cite{Chapter2012}. The smaller
particles were synthesized by co-precipitation of Fe$^{3+}$ and
Fe$^{2+}$ in oleic acid micelles. The synthesis of the longer iron
oxide particles utilized precipitation of Fe$^{2+}$ in the presence
of oleic acid as stabilizer. The morphology and size distribution of
the prepared NPs were determined by transmission electron
microscopy (TEM). Fig.~\ref{TEM} shows the TEM images of the
prepared spherical and rod-like nanoparticles. The mean diameter of the
spherical magnetic NPs was $D=10$~nm. The average diameter of the
shorter rod-like nanoparticles was $D=8$~nm and the mean length was
$L=40$~nm. The longer rod-like NPs had the average diameter of
$D=11$~nm and their mean length was $L=240$~nm.

The ferronematic samples were based on the thermotropic nematic
6CHBT which is a low-temperature-melting enantiotropic liquid crystal with high
chemical stability \cite{Dab}. The phase transition temperature from
the isotropic liquid to the nematic phase (the clearing point) of the
studied nematic was found by polarizing optical microscopy at T$_{IN}$ = 43.3$^\circ$C. The doping
was done by adding nanoparticles (spherical or rod-like) to the
liquid crystal in the isotropic phase under continuous stirring. The
nanoparticles were coated with oleic acid as a surfactant to
suppress their aggregation. The ferronematic samples were prepared
with three  different volume concentrations of the spherical as well as
of the rod-like magnetic particles: $\phi_1$ = 1 $\times$ 10$^{-5}$,
$\phi_2$ = 5 $\times$ 10$^{-5}$ and $\phi_3$ = 1 $\times$ 10$^{-4}$.

The structural transition from the isotropic to the nematic phase
was monitored by polarizing microscope as well as by capacitance
measurements. The prepared samples were filled into a capacitor made
of ITO-coated glass electrodes with the electrode area approximately
1~cm~$\times$~1~cm. The distance between the electrodes (the sample
thickness) was $d$~=~5~$\mu$m. The samples (the undoped 6CHBT, or
6CHBT doped by various NPs) were filled into the cells in the isotropic phase due
to capillary forces. A rubbed polyimide coating on the electrodes
ensured planar orientation.

For all samples the measurements started with polarizing microscopic
observations. The samples were put into a Linkam hot stage, heated
to the isotropic state, then the samples were slowly cooled (at the
rate of 1$^{\circ}$C/min) to the nematic state while monitoring
their textures between crossed polarizers. The transition
temperature T$_{IN}$ was taken as the temperature, where nematic
droplets appeared in the isotropic melt in the cooling process.
Next, the temperature dependence of the capacitance of the same
samples was measured in the same cells at the frequency of 1~kHz by
a high precision capacitance bridge Andeen Hagerling (the accuracy
at 1~kHz is 0.8~aF). The samples were first heated to the isotropic
phase and then they were again slowly cooled to the nematic phase
(at the rate of 1$^{\circ}$C/min).

\section{Results  and Discussion}

\begin{figure}[h]
\begin{center}
\includegraphics[width=30pc]{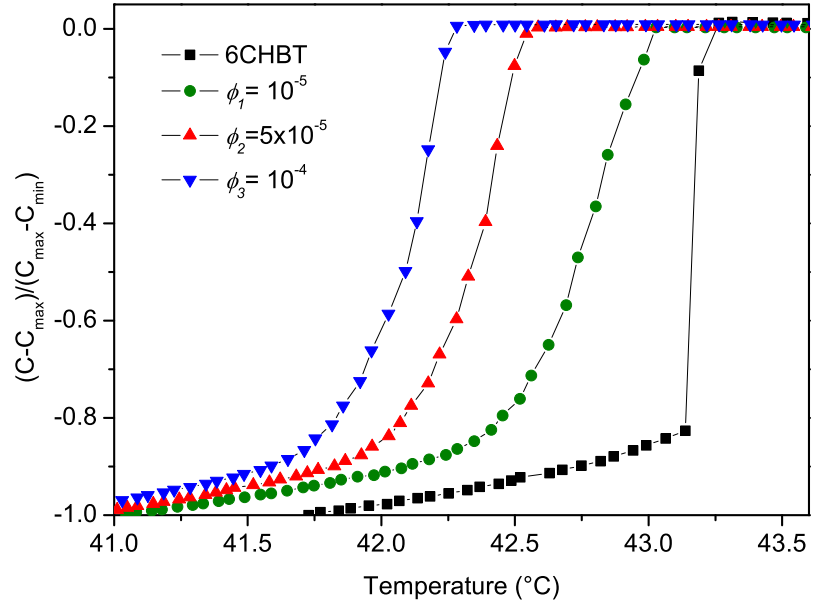}
\caption{Reduced capacitance vs. temperature for undoped 6CHBT and
for 6CHBT doped with spherical magnetic nanoparticles of different
concentrations $\phi$.} \label{spher}
\end{center}
\end{figure}

\begin{figure}[h]
\begin{center}
\includegraphics[width=30pc]{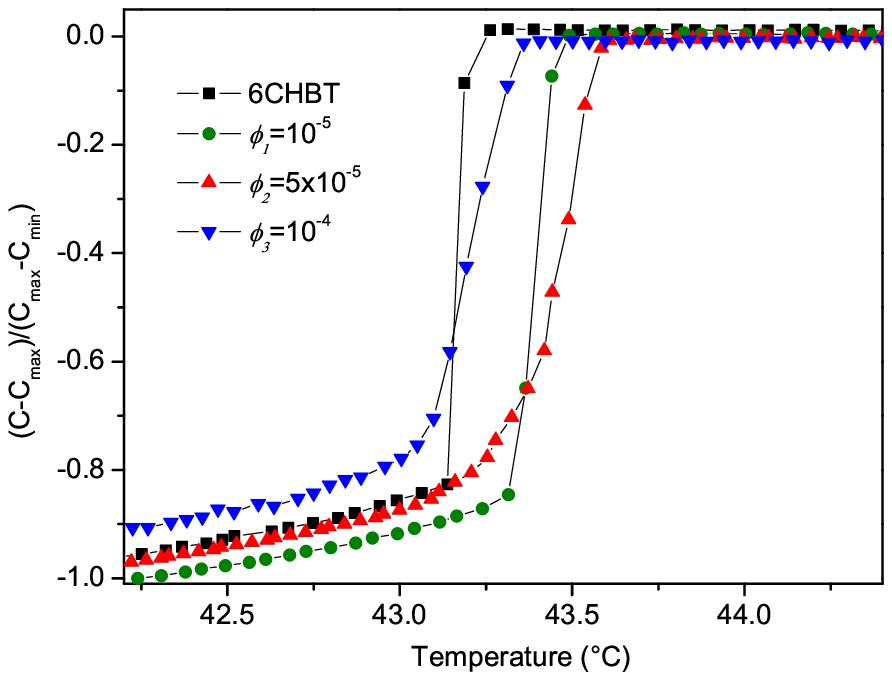}
\caption{Reduced capacitance vs. temperature for undoped 6CHBT and
for 6CHBT doped with short rod-like magnetic nanoparticles of
different concentrations $\phi$.} \label{short}
\end{center}
\end{figure}

\begin{figure}[h]
\begin{center}
\includegraphics[width=30pc]{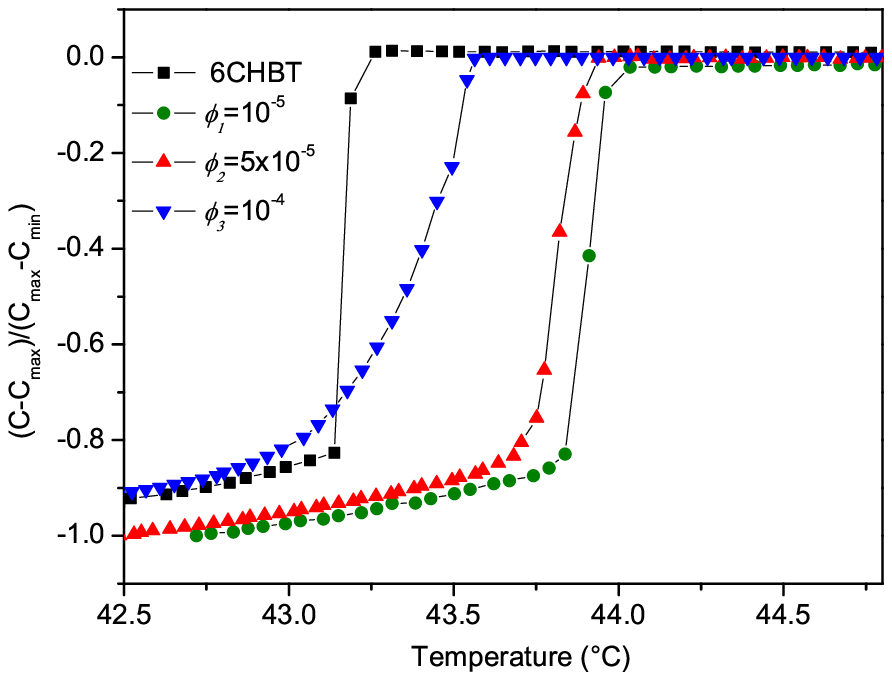}
\caption{Reduced capacitance vs. temperature for undoped 6CHBT and
for 6CHBT doped with long rod-like magnetic nanoparticles of
different concentrations $\phi$.} \label{long}
\end{center}
\end{figure}

 Fig.~\ref{spher} shows the temperature
dependence of the reduced capacitance
($C-C_{max})/(C_{max}-C_{min})$ for the undoped 6CHBT and for the
ferronematics containing spherical magnetic NPs in different volume
concentrations. Here $C$, $C_{max}$ and $C_{min}$ correspond to the
capacitances at the actual, at the highest and at the lowest
temperatures, respectively. Each curve exhibits a monotonic decrease
of the capacitance with diminishing temperature. In the isotropic
phase the capacitance is nearly constant. The sudden drop of the
capacitance indicates the appearance of the orientational order (and
hence the increase of the anisotropy), i.e. the phase transition to
the nematic phase. The decrease of $C$ occurs due to the planar
alignment and the positive dielectric anisotropy of 6CHBT and the
6CHBT-based ferronematics. Fig.~\ref{spher} clearly shows that
doping with spherical NPs results in a shift of $T_{IN}$ towards
lower temperatures; the shift becomes larger with increasing the
volume concentration of magnetic NPs. This is in accordance with the
expectations, as NPs behave as impurities introducing disorder and
thus reducing $T_{IN}$. Figs.~\ref{short} and ~\ref{long} also show
the temperature dependence of the reduced capacitance, however, for
ferronematics doped by short and long rod-like magnetic
nanoparticles, respectively. The monotonicity of the temperature
dependence and the easy detectability of the isotropic-nematic phase
transition temperature hold here too. However, in the case of doping
6CHBT with rod-like magnetic NPs, $T_{IN}$ shifts towards higher
temperatures. Moreover, though the increasing volume concentration
of NPs initially enhances the shift, this tendency turns over for
higher concentration. Nevertheless, $T_{IN}$ still remains higher
for the samples doped with rod-like NPs than that in the undoped LC.

\begin{figure}[h]
\begin{center}
\includegraphics[width=30pc]{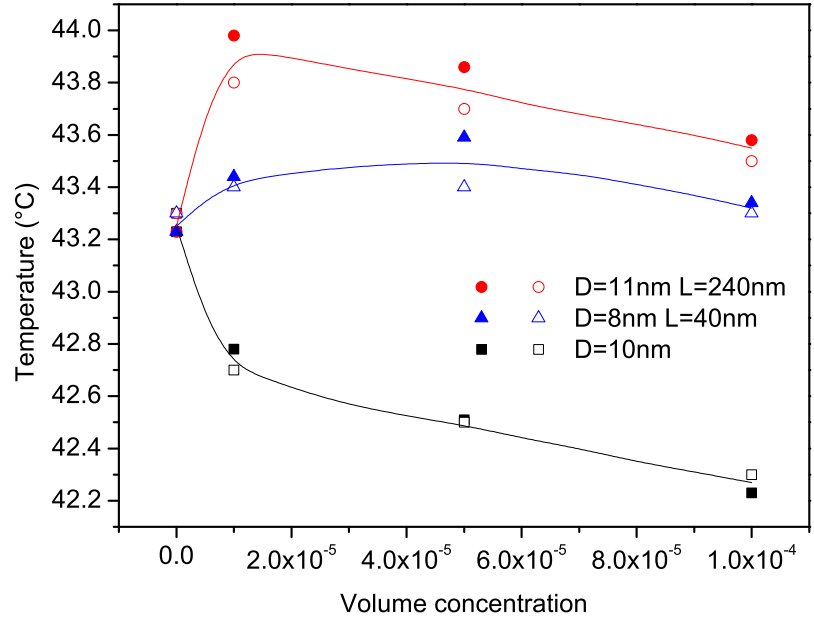}
\caption{ Dependence of the isotropic to nematic phase transition
temperature on the volume concentration for spherical  and rod-like
magnetic nanoparticles (as indicated in the legend), obtained by
capacitance measurements (full symbols) and determined by polarizing
microscopy (open symbols). The solid lines are guides to the eye.}
\label{spolu}
\end{center}
\end{figure}

The differences between the behaviour of ferronematics containing
NPs of different shape is even more perceptible in 
Fig.~\ref{spolu} which shows the dependence of the $T_{IN}$ on the
volume concentration of NPs, obtained by capacitance measurements
and by polarizing microscopy. As one can see, the
results provided by the two independent techniques are in good
agreement. Fig.~\ref{spolu} clearly shows, that the increase
of the phase transition temperature is more pronounced when long
rod-like NPs are used for the doping instead of the short ones. In
order to understand these features one has to recall the recent
theory of Gorkunov and Osipov \cite{GO}. They have developed a
mean-field molecular description of the thermodynamics of nematic
liquid crystals mixed with NPs, which may be anisotropic due to
their shape, their surface treatment, and/or their spontaneous
polarization. As it has been shown, the large shape anisotropy of
NPs (namely the rod-likeness) enhances the nematic order and, as a
consequence, increases the phase transition temperature. In contrast
to that, spherical NPs reduce the orientational order and thus lower
$T_{IN}$. These conclusions are in agreement with our experimental
findings presented in Figs.~\ref{spher}~-~\ref{spolu} and
explain the different signs of the shift of $T_{IN}$ for spherical
and rod-like NPs. Moreover, it is evident from our TEM measurements
that longer rod-like nanoparticles have a larger length/diameter
ratio ($\approx$~22) than the shorter NPs ($\approx$~5); i.e. longer
NPs have higher anisotropy and thus larger shift in $T_{IN}$.

In order to prevent aggregation of NPs, they are usually coated with
surfactants during the preparation. In a recent work various ligands
were used for this purpose for two different NPs of spherical shape
dispersed in a polymorphic LC \cite{Prodanov2012}. A slight, though measurable
change in the phase transition temperatures has been reported in these dispersions.
However, the influence of the type and concentration of the NPs, as well as the
type of the surfactant on the phase transition temperature shift
could not be distinguished clearly from these measurements.
Therefore, we have coated all magnetic NPs (spherical
and rod-like) with the same surfactant (oleic acid). The theoretical
description in \cite{GO} remains valid if the surface of the
nanoparticles is coated by organic molecules. However, one has to
take into account that in general, the surfactant dilutes the liquid
crystal, reduces its order and thus lowers the isotropic-nematic
phase transition temperature in itself. In the case of doping with
coated magnetic nanoparticles, the increasing volume concentration
of NPs increases the concentration of the surfactant as well. When
spherical NPs are used as dopants, the surfactant and the NPs have a
similar effect on $T_{IN}$ resulting in its decrease, as shown in
Fig.~\ref{spher}. In contrast to that, using rod-like dopants the
NPs and the surfactant counteract; the effect of the surfactant
(which decreases $T_{IN}$) seems to prevail at higher
concentrations. This may explain the nonmonotonic concentration
dependence of $T_{IN}$ seen in Fig.~\ref{spolu}, and is also
in good qualitative agreement with the Gorkunov-Osipov theory.

Finally, we note that in ferronematics based on 6CHBT, the magnetic moment
(which is in case of rod-like NPs parallel with their long axis) coincides
with the nematic director as it has been shown in our previous work \cite{PK1}.
However, because of the low volume concentration
(order of 10$^{-5}$-10$^{-4}$) and
small value of the magnetic moment (order of 10$^{-24}$~Tm$^3$) of NPs their magnetic
interaction can be considered negligible, and therefore, does not influence the LC order significantly.

\section{Conclusion}

We have found that the shape as well as the volume concentration of
magnetic nanoparticles have a significant influence on the
temperature of the isotropic to nematic phase transition of
ferronematics. We have justified that ferronematics doped with
rod-like magnetic nanoparticles have higher $T_{IN}$ than the host
nematic or ferronematics containing spherical nanoparticles. We have
found that doping with rod-like NPs yields a nonmonotonic
concentration dependence of $T_{IN}$, which could be
attributed to the competing influence of the nanoparticles and of
the organic surfactants they are coated with. Our results provide a
firm experimental proof for the main conclusion of the recent
mean-field theory of Gorkunov and Osipov \cite{GO}.

\section*{Acknowledgments}
This work was supported by project VEGA 0045, the Slovak Research
and Development Agency under the contract No. APVV-0171-10, Ministry
of Education Agency for Structural Funds of EU in frame of projects
6220120021 and 6220120033,  the Hungarian Research Fund OTKA K81250
and M-era.Net project MACOSYS (OTKA NN110672).

\

\end{document}